\documentclass[aps, prx, 10pt, twocolumn, superscriptaddress, longbibliography]{revtex4-2}

\usepackage{graphicx}
\usepackage{mathtools, amssymb, physics, amsthm, bm, bbm}
\usepackage{tabularx, array}
\usepackage{hyperref}
\hypersetup{
 pdftitle = {Concentration of ergotropy},
 pdfauthor = {K. Hovhannisyan, R. Simon, J. Anders},
 breaklinks=true,
 pdfnewwindow=true,
 colorlinks=true,
 linkcolor=red,
 citecolor=blue,
 filecolor=blue,
 urlcolor=blue
}
\usepackage{xcolor}
\usepackage{psfrag}
\usepackage{soul, yfonts, textcomp, pifont}
\usepackage[utf8]{inputenc}
\usepackage[english]{babel}
\usepackage[T1]{fontenc}
\usepackage{dcolumn} 
\usepackage{tikz}
\usetikzlibrary{quantikz2}

\newtheorem{theorem}{Theorem}
\newtheorem{corollary}{Corollary}[theorem]

\let\tr\relax
\DeclareMathOperator{\tr}{Tr}
\let\var\relax
\DeclareMathOperator{\var}{Var}

\DeclareMathOperator{\prob}{Prob}
\DeclareMathOperator{\spr}{spr}

\newcommand{\av}[1]{\langle #1 \rangle}
\newcommand{\bav}[1]{\big\langle #1 \big\rangle}

\newcommand{\id}{\mathbbm{1}}

\newcommand{\RN}[1]{\textup{\uppercase\expandafter{\romannumeral#1}}}

\newcommand{\erg}{\mathcal{E}}

\renewcommand{\P}{\mathcal{P}}
\newcommand{\T}{\mathcal{T}}

\newcommand{\DBu}{\mathcal{D}_{\mathrm{B}}}
\newcommand{\DTr}{\mathcal{D}_{\mathrm{Tr}}}
\newcommand{\DHS}{\mathcal{D}_{\mathrm{HS}}}
\newcommand{\opn}[1]{\Vert #1 \Vert_{\mathrm{op}}}

\newcommand{\avHS}[1]{\big\langle #1 \big\rangle_{\mathrm{HS}}}
\newcommand{\avBu}[1]{\big\langle #1 \big\rangle_{\mathrm{B}}}

\begin{document}

\title{Concentration of ergotropy in quantum many-body systems}

\author{Karen Hovhannisyan}
\email{karen.hovhannisyan@uni-potsdam.de}
\affiliation{University of Potsdam, Institute of Physics and Astronomy, Karl-Liebknecht-Str. 24-25, 14476 Potsdam, Germany}

\author{Rick P. A. Simon}
\affiliation{Department of Physics and Astronomy, University of Exeter, Stocker Road, Exeter EX4 4QL, UK}

\author{Janet Anders}
\affiliation{University of Potsdam, Institute of Physics and Astronomy, Karl-Liebknecht-Str. 24-25, 14476 Potsdam, Germany}
\affiliation{Department of Physics and Astronomy, University of Exeter, Stocker Road, Exeter EX4 4QL, UK}

\begin{abstract}

Ergotropy---the maximal amount of unitarily extractable work---quantifies the ``charge level'' of a quantum battery. We prove that, in large many-body batteries, it exhibits a concentration of measure phenomenon. Namely, the ergotropy almost coincides with its average for almost all states sampled from the Hilbert--Schmidt measure. Importantly, the average ergotropy is macroscopic, scaling with the number of particles in the system. This occurs despite the von Neumann entropy also concentrating around a macroscopic average, a surprising coexistence since high entropy is commonly associated with effective thermality and thus passivity. We establish these results by proving that ergotropy, as a function of the state, is Lipschitz continuous with respect to the Bures distance and by applying Lévy's measure concentration lemma. Furthermore, we numerically extend our findings to the Bures measure, which represents the situation with the least amount of prior information about the state. Our continuity and concentration bounds open a new perspective on the noise-robustness of quantum many-body batteries and lay the groundwork for further rigorous studies of their charge stability and work extraction with limited prior knowledge.

\end{abstract}

\maketitle

\section{Introduction}
\label{sec:intro}

Ergotropy is the maximal average work one can extract from a system by means of cyclic Hamiltonian processes \cite{Allahverdyan_2004}. Along with the notion of passivity \cite{Hatsopoulos_1976, Pusz_1978, Lenard_1978} associated with it, ergotropy is a powerful tool that has provided numerous insights into the workings of thermodynamic stability and the second law of thermodynamics in general \cite{Hatsopoulos_1976, Pusz_1978, Lenard_1978, Allahverdyan_2004, Alicki_2013, Sparaciari_2017}. But more importantly from a practical perspective, it is used in quantum thermodynamics to characterize the amount of ``useful energy,'' namely, work, stored in quantum batteries \cite{Alicki_2013, Gelbwaser-Klimovsky_2013, Hovhannisyan_2013, Perarnau_2015, Binder_2015, Campaioli_2017, Ferraro_2018, Barra_2019, Farina_2019, Andolina_2019, Niedenzu_2019, Gherardini_2020, Hovhannisyan_2020, Cakmak_2020, Seah_2021, Tirone_2021, Barra_2022, Gyhm_2022, Tirone_2024, Lobejko_2022, Yang_2023, Campaioli_2024, Song_2024, Feliu_2024, Simon_2025}.

Calculating ergotropy requires reordering the eigenvalues of the state of the system \cite{Allahverdyan_2004} (see below). Therefore, due to this combinatorial aspect, ergotropy is hard to compute and, in general, analyze in large Hilbert spaces. However, in all imaginable real-life scenarios, batteries would be required to store large amounts of work, which means that they would have to be large, many-body systems. Thus, gaining a better understanding of ergotropy in large systems is not only of fundamental interest but also central to the field of quantum batteries and their (future) applications.

In this paper, we demonstrate that the ergotropy of a very general class of many-body systems exhibits measure concentration in the large-system limit. Namely, it almost coincides with its ensemble average for almost all states \cite{Talagrand_1996, Milman_book_2001} sampled from the Hilbert--Schmidt measure \cite{Bengtsson_book_2006}. We establish this by first proving that ergotropy is a Lipschitz continuous function of the state with respect to the Bures distance (as well as trace and Hilbert--Schmidt distances) in the most general setting. Then, we use this continuity bound along with Lévy's measure concentration lemma \cite{Talagrand_1996, Milman_book_2001, Watrous_book_2018} to show that ergotropy concentrates around its ensemble average for systems whose largest energy eigenvalue scales at most polynomially with the system size. This condition is fairly loose, and it comfortably encompasses ordered and disordered spin lattices with arbitrary ranges of interactions (see Sec.~\ref{sec:manybody}). As for the ensemble average of ergotropy itself, we numerically demonstrate that it is macroscopic in that it also scales polynomially with the system size.


A random-state ensemble can also be interpreted inferentially, with the measure over state space playing the role of a prior encoding our knowledge about an otherwise unknown state. In this context, the Hilbert--Schmidt measure is not distinguished in any particular way, whereas the Bures measure \cite{Slater_1996, Hall_1998} is considered to be the least informative prior over quantum states \cite{Slater_1996, Hall_1998, Slater_1998b, Bengtsson_book_2006}. This motivates us to revisit the problem above for states sampled from the Bures measure. Due to the Lipschitz continuity of ergotropy and the von Neumann entropy (see Ref.~\cite{Sekatski_2021} and Sec.~\ref{sec:entropy_lipschitz}) with respect to the Bures distance, one is naturally led to expect that both would concentrate also in this case. The difficulty is that, unlike for the Hilbert--Schmidt measure, no exponential concentration inequalities analogous to Lévy's lemma are currently known for the Bures measure \cite{Slater_2005}. Nonetheless, through extensive numerical simulations, we show in this paper that ergotropy and entropy do concentrate also in this case.

\medskip

Lastly, we remark that randomizing several battery characteristics has been investigated in Refs.~\cite{Caravelli_2020, Ghosh_2020}. However, those works studied settings and objectives distinct from those considered here, and did not consider ergotropy.

\section{Continuity of ergotropy}
\label{sec:ergotropy_lipschitz}

In this section, we prove that ergotropy, denoted by $\erg$, as a function of the state $\rho$, is Lipschitz continuous with respect to three different distances: the trace, Hilbert--Schmidt, and Bures distances. This result is fully general---no assumptions on the Hamiltonian (except for it to have a finite norm) are made here.

First of all, let us summarize the main points about ergotropy. It is the maximal amount of work that can be extracted from a system by cyclicly varying its Hamiltonian $H$. Namely,
\begin{align} \label{ergot_def_1}
    \erg(\rho, H) \coloneq \tr(\rho H) - \min\limits_U \tr(U \rho U^\dagger H),
\end{align}
where the minimization is over \textit{all} unitaries $U$, and the minimum is achieved by the unitary $U_{\rho^\downarrow}$ that transforms $\rho$ into the passive state $\rho^\downarrow$ \cite{Hatsopoulos_1976, Pusz_1978, Lenard_1978, Allahverdyan_2004}. For a system with a $d$-dimensional Hilbert space, the passive state is given by
\begin{align}
    \rho^\downarrow = \sum_{k = 0}^{d-1} \lambda_k^\downarrow \dyad{E_k},
\end{align}
where $\{\ket{E_k}\}_{k=0}^{d-1}$ is an eigenbasis of $H$ with the corresponding eigenvalues ordered increasingly, $E_0 \leq \cdots \leq E_{d-1}$, and $\lambda_0^\downarrow \geq \cdots \geq \lambda_{d-1}^\downarrow \geq 0$ are $\rho$'s eigenvalues ordered decreasingly. Thus, the ergotropy can also be written as 
\begin{align} \label{ergot_def_2}
    \erg(\rho, H) = \tr[(\rho - \rho^\downarrow) H].
\end{align}
We stress that, by design, $\erg$ only depends on $\rho$ and $H$. Note that, while $\rho^\downarrow$ is determined solely by $\rho$ and $H$, it (and therefore also $U_{\rho^\downarrow}$) will not be unique when $\rho$ or $H$ have degenerate spectra.

\medskip

Now, recall that a function $f : M \mapsto \mathbbm{R}$, where $M$ is some metric space with distance function $D : M \times M \mapsto \mathbbm{R}_+$, is called Lipschitz continuous with respect to the distance $D$, if
\begin{align*}
    \exists L > 0 \;\; \text{such that} \;\; |f(x) - f(y)| \leq L \, D(x, y) \;\; \forall x, y \in M.
\end{align*}
The constant $L$ must not depend on $x$ and $y$, and the smallest such $L$ for a given distance function $D$, denoted $L^f_D$, is called the Lipschitz constant of $f$.

The main result of this section can be summarized as the following theorem proven in Appendix~\ref{app:erg_cont} by using the Lidskii--Wielandt theorem \cite{Lidskii_1950, Mirsky_1960, Bhatia_book_1997}.

\begin{theorem} \label{thm:ergotropy_lipschitz}
    Ergotropy is a Lipschitz continuous function of the state with respect to the trace distance $(\mathrm{Tr})$, Bures distance $(\mathrm{B})$, and Hilbert--Schmidt distance $(\mathrm{HS})$. The corresponding Lipschitz constants satisfy
    \begin{align} \label{erg_lipschs}
    L_{\mathrm{B}}^\erg, \; L_{\mathrm{Tr}}^\erg \leq 2 \, \spr(H) \;\;\quad \mathrm{and} \;\;\quad L_{\mathrm{HS}}^\erg \leq \sqrt{d} \, \spr(H),
    \end{align}
    where $\spr(H) \coloneq E_{d-1} - E_0$ is the spectral spread of $H$.
\end{theorem}

Interestingly, the spectral spread of $H$ is the maximum possible ergotropy the system can host, i.e., $\max_\rho \erg(\rho, H) = \spr(H)$. Moreover, it is closely related to the operator norm $\opn{\bullet}$: $\spr(H) \leq 2 \opn{H}$ for any $H$ \cite{HornJohnson}, and whenever $E_0 = 0$, one simply has $\spr(H) = \opn{H}$.

We also remind the reader that the Bures distance is defined as \cite{Bures_1969}
\begin{align} \label{DBu_def}
    \DBu(\rho, \sigma) \coloneq \sqrt{2 - 2\,\sqrt{F(\rho, \sigma)}},
\end{align}
where $F(\rho, \sigma) = \big[\tr\big(\sqrt{\rho^{1/2} \sigma \rho^{1/2}}\big)\big]^2$ is the Uhlmann fidelity \cite{Uhlmann_1976}. Whereas the trace and Hilbert--Schmidt distances are given by, respectively,
\begin{align} \label{DTr_def}
    \DTr(\rho, \sigma) \coloneq \frac{1}{2} \Vert \rho - \sigma \Vert_1 \quad \text{and} \quad \DHS(\rho, \sigma) \coloneq \Vert \rho - \sigma \Vert_2,
\end{align}
where $\Vert A \Vert_p \coloneq (\tr[(A^\dagger A)^{p/2}])^{1/p}$ is the Schatten $p$-norm, with $A$ an arbitrary operator and $p \geq 1$.

\medskip

Finally, let us introduce an extra copy of the system's Hilbert space $\mathcal{H}_d$ and consider purifications $\ket{\P(\rho)} \in \mathcal{H}_d \otimes \mathcal{H}_d$ of states $\rho$ on $\mathcal{H}_d$. Although the purification of a given state is not unique, so $\ket{\P(\rho)}$ rather represents a set, any $\ket{\P(\rho)}$ satisfies $\tr_\mathrm{extra \, copy}[\dyad{\P(\rho)}] = \rho$. This allows us to view ergotropy as a function of purifications and study its continuity with respect to the Euclidean distance. Indeed, combining Theorem~\ref{thm:ergotropy_lipschitz} with Uhlmann's theorem \cite{Uhlmann_1976}, we obtain the following corollary, whose proof is given in Appendix~\ref{app:erg_cont}.
\begin{corollary} \label{thm:ergotropy_lipschitz_purified}
Ergotropy is Lipschitz continuous as a function of $\ket{\P(\rho)}$ with respect to the Euclidean distance:
\begin{align} \label{erg_lips_euclid}
    \vert \erg(\rho, H) - \erg(\sigma, H)| \leq 2 \spr(H) \, \Vert \! \ket{\P(\rho)} - \ket{\P(\sigma)} \! \Vert_2 \, ,
\end{align}
where $\Vert \! \ket{\phi} \! \Vert_2 = \sqrt{\braket{\phi}{\phi}}$ is the Euclidean norm. Thus, the corresponding Lipschitz constant of $\erg$ is
\begin{align} \label{koxac1}
    L_{\mathrm{E}}^\erg \leq 2 \spr(H).
\end{align}
\end{corollary}

\section{Continuity of entropy}
\label{sec:entropy_lipschitz}

Alongside ergotropy, it is insightful to also consider entropy. It is known \textit{not} to be Lipschitz continuous with respect to trace distance. Indeed, the standard continuity inequality for entropy \cite{Fannes_1973, Audenaert_2007, Petz_book_2008} is
\begin{align} \label{entropy_cont}
|S(\rho) - S(\sigma)| \leq \DTr(\rho, \sigma) \, \ln d + h\big(\DTr(\rho, \sigma)\big),
\end{align}
where $h(x) = - x \ln x - (1-x) \ln(1-x)$ is the binary entropy. Hence, since the derivative of $h(x)$ at $x = 0$ diverges logarithmically, $\nexists C$ such that $h(x) \leq C x$. Note that this means that $S(\rho)$ is \textit{not} Lipschitz continuous with respect to the Hilbert--Schmidt distance as well, since in finite dimensions the trace and Hilbert--Schmidt distances are norm-equivalent.

\medskip

Entropy is however Lipschitz continuous with respect to the Bures distance. This is a straightforward consequence of Lipschitz-continuity of $S(\rho)$ with respect to the so-called Bures angle proven in Ref.~\cite{Sekatski_2021}. We defer the details of this simple demonstration to Appendix~\ref{app:ent_cont}, summarizing here that
\begin{align} \label{entropy_lips_Bu}
    |S(\rho) - S(\sigma)| \leq L_{\mathrm{B}}^S \, \DBu(\rho, \sigma),
\end{align}
where, for $d \geq 5$,
\begin{align} \label{entropy_lips_Bu_coeff}
L_{\mathrm{B}}^S \leq \frac{\pi \ln d}{\ln 2}.
\end{align}
Smaller $d$'s are irrelevant for us as we are primarily interested in large systems.

As in Sec.~\ref{sec:ergotropy_lipschitz}, employing Uhlmann's theorem [specifically, its direct consequence, Eq.~\eqref{bures-leq-euclid}] yields
\begin{align} \label{hayden_ineq}
    |S(\rho) - S(\sigma)| \leq L_{\mathrm{E}}^S \, \Vert \ket{\P(\rho)} - \ket{\P(\sigma)} \! \Vert_2\, ,
\end{align}
where, as above, $\ket{\P(\rho)}$ is the purification of $\rho$ on $\mathcal{H}_d \otimes \mathcal{H}_d$ and $L_{\mathrm{E}}^S$---the Lipschitz constant of $S$ as a function of $\ket{\P(\rho)}$ with respect to the Euclidean norm---is $\leq L_{\mathrm{B}}^S$. This bound on $L_{\mathrm{E}}^S$ is however slightly cruder than that proven in Ref.~\cite[Lemma III.2]{Hayden_2006}, where it was shown that
\begin{align} \label{koxac2}
    L_{\mathrm{E}}^S \leq \frac{\sqrt{8} \ln d}{\ln 2}.
\end{align}

\section{Concentration of ergotropy and entropy}
\label{sec:concentration}

The principle of measure concentration roughly states that a random variable that is a Lipschitz continuous function of many random variables is essentially constant \cite{Talagrand_1996, Milman_book_2001}. Namely, its distribution is concentrated around its median (therefore, also the mean). Here, using the continuity bounds in Eqs.~\eqref{erg_lips_euclid} and~\eqref{hayden_ineq}, we will apply Lévy's measure concentration lemma \cite{Talagrand_1996, Milman_book_2001, Watrous_book_2018} to show that, in high dimensions, $\erg(\rho)$ and $S(\rho)$ concentrate around their respective ensemble averages when $\rho$ is sampled from the Hilbert--Schmidt measure.

We begin by stating Lévy's lemma. Let $f(x)$ be a function from the $(n-1)$-dimensional unit sphere $\mathbbm{S}^{n-1}$ to the real line $\mathbbm{R}^1$. Furthermore, let $f(x)$ be Lipschitz continuous with respect to the Euclidean distance in $\mathbbm{R}^n$---the $n$-dimensional flat space that contains $\mathbbm{S}^{n-1}$. According to Lévy's lemma, if $x$ is distributed uniformly on $\mathbbm{S}^{n-1}$ with respect to the Haar probability measure, then the probability that $f(x)$ deviates from its mean by more than $\ell$ satisfies the bound
\begin{align} \label{Levis}
    \prob\big[ |f(x) - \av{f(x)}| > \ell \big] \leq 3 \, e^{- \ell^2 / \Upsilon_f^2}.
\end{align}
The average $\av{f(x)}$ is over the Haar measure on the sphere, and
\begin{align}
    \Upsilon_f = \frac{L_\mathrm{E}^f}{\sqrt{\alpha n}},
\end{align}
where $L_\mathrm{E}^f$ is the Lipschitz constant of $f(x)$ and $\alpha > 0$ is a universal (i.e., $f$-independent) constant. 

The width of the distribution of $f(x)$ around $\av{f(x)}$ is governed by $\Upsilon_f$, which can therefore be considered an analogue of the standard deviation of $f(x)$. To get a tighter bound in Eq.~\eqref{Levis}, $\alpha$ should be chosen as large as possible (note that, if Eq.~\eqref{Levis} holds for some $\alpha$, then it automatically holds for $\forall \alpha' \leq \alpha$). The largest value of $\alpha$ we found in the literature is in Ref.~\cite[Theorem 7.37]{Watrous_book_2018}, where the bound in Eq.~\eqref{Levis} is proven for $\alpha = 1 / (25 \pi)$. We will henceforth use this value for $\alpha$. 

\medskip

Let us now apply Lévy's lemma to the ergotropy and entropy. To that end, we first introduce the partial tracing map $\T : \mathcal{H}_d \otimes \mathcal{H}_d \to \mathcal{H}_d$,
\begin{align} \label{parttracemap}
    \T(\ket{\phi}) \coloneq \tr_{\substack{\mathrm{extra \, copy} \\ \mathrm{of} \, \mathcal{H}_d}} \big[ \dyad{\phi} \big],
\end{align}
such that $\ket{\phi} \in \ket{\P(\T(\ket{\phi}))}$. As was proven in Ref.~\cite{Zyczkowski_2001}, sampling $\ket{\psi}$ from the Fubini--Study measure on the pure states in $\mathcal{H}_d \otimes \mathcal{H}_d$ exactly corresponds to sampling states $\T(\ket{\psi})$ from the Hilbert--Schmidt measure in the Hilbert space $\mathcal{H}_d$. In its turn, the Fubini--Study measure on the pure states in $\mathcal{H}_d \otimes \mathcal{H}_d$ corresponds to the Haar measure on $\mathbbm{S}^{2 d^2 - 1}$ \cite{Bengtsson_book_2006}. By employing this chain of correspondences and invoking the continuity bounds~\eqref{erg_lips_euclid} and~\eqref{hayden_ineq}, we show in Appendix~\ref{app:hilbert-to-sphere} that Lévy's lemma leads to the following concentration inequalities for ergotropy and entropy. Specifically, for ergotropy we have that
\begin{align} \label{conc_erg}
    \prob\Big[ \big| \erg(\rho, H) - \avHS{\erg(\rho, H)} \big| > \ell \Big] \leq 3 \, e^{- \ell^2 / \Upsilon_{\erg}^2} \, ,
\end{align}
with
\begin{align} \label{width_erg}
    \Upsilon_{\erg} = \frac{\sqrt{25 \pi} L_{\mathrm{E}}^\erg}{\sqrt{2 d^2}} \leq \frac{\sqrt{50 \pi} \, \spr(H)}{d}.
\end{align}
Likewise, for entropy,
\begin{align} \label{conc_ent}
    \prob\Big[ \big| S(\rho) - \avHS{S(\rho)} \big| > \ell \Big] \leq 3 \, e^{- \ell^2 / \Upsilon_{S}^2} \, ,
\end{align}
with
\begin{align} \label{width_ent}
    \Upsilon_{S} = \frac{\sqrt{25 \pi} L_{S}^{\mathrm{E}}}{\sqrt{2 d^2}} \leq \frac{\sqrt{100 \pi} \, \ln d}{d \ln 2}.
\end{align}
We see that, for $d \gg 1$, $\Upsilon_{S} \ll 1$, indicating exponential concentration of the distribution of $S$ around its ensemble average. It is worth noting here that, curiously, $S(\rho)$ concentrates under the Hilbert--Schmidt measure despite not being Lipschitz continuous with respect to the Hilbert--Schmidt distance \cite{Hayden_2006}.

\medskip

In contrast to entropy, the concentration properties of ergotropy strongly depend on $\spr(H)$. Indeed, if, e.g., $\spr(H) \propto d$, then, while the distribution of $\erg$ with respect to $\ell$ will still have exponentially decaying tails, it will not necessarily concentrate strongly around its average for large values of $d$. To single out the dependence of the statistics of the ergotropy on $\spr(H)$ and remove the units of energy from ergotropy, let us introduce the dimensionless normalized Hamiltonian
\begin{align} \label{hhat_def}
    \widehat{H} \coloneq \frac{H}{\spr(H)}.
\end{align}
Reading directly from Eq.~\eqref{erg_lips_euclid}, the Lipschitz constant of $\erg\big(\rho, \widehat{H}\big)$ with respect to the Euclidean distance in the purified space, $\widehat{L}_{\mathrm{E}}^\erg \leq 2$. Accordingly, Eqs.~\eqref{conc_erg} and~\eqref{width_erg} rewrite as
\begin{align} \label{conc_wterg}
    \prob\Big[ \big| \erg\big(\rho, \widehat{H}\big) - \avHS{\widehat{\erg}} \big| > \ell \Big] \leq 3 \, e^{- \ell^2 / \widehat{\Upsilon}_{\erg}^2} \, ,
\end{align}
with
\begin{align} \label{width_erghat_1}
    \widehat{\Upsilon}_{\erg} \leq \frac{\sqrt{50 \pi}}{d},
\end{align}
where we, for convenience, denoted
\begin{align} \label{Erg_hat}
    \avHS{\widehat{\erg}} \coloneq \avHS{\erg\big(\rho, \widehat{H}\big)}.
\end{align}
Switching to $\widehat{H}$ thus removes the inconvenient dependence of the average and distribution width of the ergotropy on the dimensionful parameter $\spr(H)$. Reverting to $\erg(\rho, H)$ is straightforward:
\begin{align} \label{hat_to_original}
\begin{split}
    \Upsilon_{\erg} &= \spr(H) \, \widehat{\Upsilon}_{\erg},
    \\
    \avHS{\erg(\rho, H)} &= \spr(H) \, \avHS{\widehat{\erg}}.
\end{split}
\end{align}

\section{Implications for many-body systems}
\label{sec:manybody}

\begin{figure}[t!]
    \centering
    \includegraphics[width=0.95\linewidth]{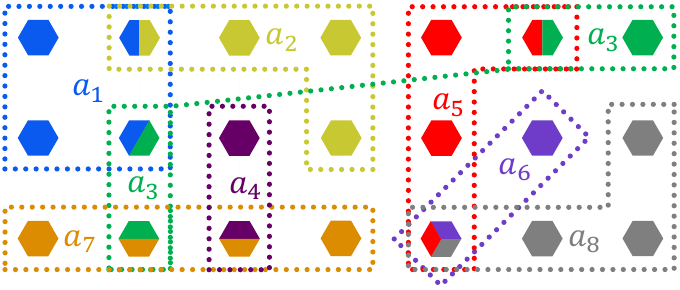}
    \caption{Illustration of the general structure of an $r$-body lattice Hamiltonian. Each dotted enclosure $a$ delineates the subset of vertices coupled by the corresponding $h_a$. The subsets are also color-coded to further facilitate visual identification. In this example, $r = 4$ and $a$ runs from $a_1$ to $a_8$. The subset $a_3$ (green), consisting of two far-apart vertex pairs, exemplifies a long-range interaction, with $h_{a_3}$ coupling disconnected regions of the lattice.}
    \label{fig:many_body_ham}
\end{figure}

A natural class of large systems exhibiting concentration of ergotropy---namely, $\Upsilon_\erg \to 0$ as $d \to \infty$---are general many-body systems living in finite-dimensional Hilbert spaces. To see that, let us first define what we mean by a general many-body system. Suppose the system consists of $N \gg 1$ components such that the local Hilbert-space dimension does not vary too much from component to component. This guarantees that $\ln d \propto N$, and more specifically, that there exists a constant $\kappa > 0$ such that
\begin{align} \label{N-to-d}
    d \geq e^{\kappa N}.
\end{align}
Furthermore, we assume that the system features at most $r$-body interactions for some arbitrary but fixed and $N$-independent $r$ ($N \gg r$). Namely, the Hamiltonian of the whole system is of the form
\begin{align} \label{ham_gen}
    H = \sum_a h_a,
\end{align}
where $a$ labels the local interaction terms $h_a$ each of which acts nontrivially on at most $r$ components (see Fig.~\ref{fig:many_body_ham} for a sketch). We moreover assume that $\opn{h_a} \leq \mathfrak{h}$ $\forall a$ for some $N$-independent constant $\mathfrak{h} > 0$. This construction is fairly general and includes all conceivable ordered or disordered spin-lattice-type systems in any dimension and geometry and featuring local interactions of arbitrary range \cite{Gogolin_2016, Mori_2018}.

Given that there are at most $\binom{N}{r}$ terms in the sum in Eq.~\eqref{ham_gen}, and that, for $N \gg r$, $\binom{N}{r} \propto N^r$, the triangle inequality of the operator norm immediately yields $\opn{H} = O\big[(\ln d)^r\big]$.
In what follows, we will use the following equivalent form of this fact:
\begin{align} \label{ham_scal_gen}
    \spr(H) \leq E \, (\ln d)^r \, \propto E \, N^r,
\end{align}
where $E$ has units of energy and is \textit{independent} of $N$. This quantity can be viewed as an (averaged) energy scale characterizing the individual local Hamiltonian terms $h_a$.

Now, Eq.~\eqref{N-to-d} alone is already sufficient to conclude that the concentration of $S(\rho)$ and $\erg\big(\rho, \widehat{H}\big)$ is exponential not only in the sense of exponentially decaying tails [Eqs.~\eqref{conc_ent} and~\eqref{conc_wterg}] but also in the sense that the widths of their distributions around their averages is exponentially small in $N$. Indeed, plugging Eq.~\eqref{N-to-d} into Eqs.~\eqref{width_ent} and~\eqref{width_erghat_1}, we get
\begin{align}
    \Upsilon_S = O\Big( \frac{N}{e^{\kappa N}}\!\Big) \quad \mathrm{and} \quad \widehat{\Upsilon}_\erg = O\Big( \frac{1}{e^{\kappa N}}\!\Big).
\end{align}
Moreover, since for our very generally and inclusively defined local Hamiltonians $\spr(H)$ can scale only polynomially with $N$ [Eq.~\eqref{ham_scal_gen}], it immediately follows from substituting it into Eq.~\eqref{hat_to_original} [or directly into Eq.~\eqref{width_erg}] that
\begin{align} \label{conc_erg_width}
    \frac{\Upsilon_{\erg}}{E} = O\Big(\frac{N^r}{e^{\kappa N}}\!\Big).
\end{align}
Namely, the ergotropy of such systems always concentrates exponentially, in both senses. We emphasize that this holds regardless of the strength and range of interactions within the system.

\medskip

Having established that ergotropy and entropy concentrate around their averages, let us now look at the averages $\av{\erg(\rho, H)}_{\mathrm{HS}}$ and $\av{S(\rho)}_{\mathrm{HS}}$ themselves. The average of the entropy is well-known \cite{Hayden_2006} and is
\begin{align} \label{avent_scaling}
    \avHS{S(\rho)} \, \propto \, \ln d \, \propto \, N,
\end{align}
as expected.

To gain insight into the average of ergotropy, let us, for intuitive clarity and without loss of any generality, shift $\widehat{H}$ by a constant such that its smallest eigenvalue is $0$. Now, to derive an upper bound on $\avHS{\widehat{\erg}}$, we invoke Eq.~\eqref{ergot_def_2} to obtain
\begin{align*}
    \avHS{\widehat{\erg}} = \tr\big[\avHS{\rho} \widehat{H}\big] - \tr\big[\avHS{\rho^\downarrow} \widehat{H}\big].
\end{align*}
Due to the unitary-invariance of the Hilbert--Schmidt measure \cite{Bengtsson_book_2006}, the average state $\avHS{\rho}$ must also be unitary-invariant. Since the only such operator (up to a constant factor) is $\id$, we have that $\avHS{\rho} = \id / d$. Namely, the average state is the infinite-temperature Gibbs state. Thus, taking into account that $\tr\big[\avHS{\rho^\downarrow} \widehat{H}\big] \geq 0$,
\begin{align} \label{av_bound_1}
    \avHS{\widehat{\erg}} \leq \frac{1}{d} \tr \widehat{H} < 1.
\end{align}
Note that, since the initial average energy $\tr(\rho \widehat{H})$ is Lipschitz continuous with respect to $\rho$, with Lipschitz constant being $\leq 1$ [see Eq.~\eqref{erg_lip_1}], the values of $\tr(\rho \widehat{H})$ are exponentially concentrated around $\tr \widehat{H} / d$. Namely, when $N \gg 1$, $\tr(\rho \widehat{H}) \approx \tr \widehat{H} / d$ for almost all $\rho$ sampled from the Hilbert--Schmidt measure.

For a lower bound, we have the obvious $\avHS{\widehat{\erg}} > 0$ stemming from the fact that ergotropy is a nonnegative quantity and is strictly positive for some states. While true for any finite $N$, this does not yet guarantee that $\avHS{\widehat{\erg}}$ tends to a strictly positive value in the $N \to \infty$ limit, i.e., that
\begin{align} \label{avene_1}
    \exists \epsilon > 0 \quad \mathrm{s.t.} \quad \avHS{\widehat{\erg}} \geq \epsilon \quad \forall d.
\end{align}
A necessary prerequisite for this hypothesis to be true is for the upper bound in Eq.~\eqref{av_bound_1} to remain finite as the system size increases. Namely,
\begin{align} \label{Ham_cond}
    \frac{1}{d} \tr\widehat{H} \geq \hat{\mathfrak{h}} \quad \forall d
\end{align}
must hold for some $d$-independent $\hat{\mathfrak{h}} > 0$. This condition is natural for many-body systems since it simply stipulates that, going back to $H$ [Eq.~\eqref{hat_to_original}], the system has a macroscopic energy when its temperature is infinite. We henceforth assume that the Hamiltonian satisfies the condition in Eq.~\eqref{Ham_cond}.

Of course, the condition in Eq.~\eqref{Ham_cond} alone does not necessitate Eq.~\eqref{avene_1}. However, there are good reasons to expect that it is nonetheless true. Indeed, the set of states with population inversions, which is when there exist at least two levels $E_k < E_j$ such that $\bra{E_k} \rho \ket{E_k} < \bra{E_j} \rho \ket{E_j}$, has strictly positive measure, and each such state has a positive ergotropy. Moreover, the set of diagonal states has measure zero, so almost all states are coherent in the energy eigenbasis, and whenever these coherences connect different energy eigensubspaces, the ergotropy is positive. The latter is true provided the Hamiltonian is not exceptionally degenerate, as is generically the case.

While the above arguments make a strong intuitive case for Eq.~\eqref{avene_1}, proving it rigorously is considerably more difficult. We therefore resort to numerical simulations to demonstrate that it indeed holds. To that end, for each $d$, we sample a large number of random states from the Hilbert--Schmidt measure and calculate the ergotropy of each state. An efficient way of generating Hilbert--Schmidt-random density matrices is provided by a result proven in Ref.~\cite{Zyczkowski_2001} stating that the states
\begin{align} \label{HS_rs}
    \rho = \frac{G G^\dagger}{\tr (G G^\dagger)},
\end{align}
where $G$ are $d\times d$ random matrices sampled from the Ginibre ensemble, are distributed according to the Hilbert--Schmidt measure. Recall that the elements of random Ginibre matrices are independent, identically distributed Gaussian random variables with mean $=0$ and variance $=1$ \cite{Mezzadri_2007}.

To eliminate any possible dependence on accidental features of the Hamiltonian, we also sample the Hamiltonians randomly, choosing the Gaussian Unitary Ensemble (GUE) \cite{Edelman_2005} as a natural measure. The GUE is unitary invariant, as required in the absence of a preferred basis, and is maximally random in the sense that all $d^2$ independent real parameters of the Hamiltonian are statistically independent. Moreover, being a product of Gaussian measures, the GUE exhibits concentration of measure \cite{Milman_book_2001, Ledoux_book_2001}. Thus, since $\tr H$ is Lipschitz continuous in $H$, it is close to $\av{\tr H}_{\mathrm{GUE}}$ for almost all GUE realizations for $d \gg 1$. Crucially, this means that although $H$ itself is random, the bound~\eqref{av_bound_1} and condition~\eqref{Ham_cond} basically remain fixed. Lastly, GUE matrices can conveniently be generated \cite{Edelman_2005} by sampling a Ginibre $d\times d$ matrix $G$ as above and forming $(G+G^\dagger)/2$. The final tweak needed for sampling $\widehat{H}$ [Eq.~\eqref{hhat_def}], is generating a $(G + G^\dagger)/2$, diagonalizing it, shifting the spectrum so that the smallest eigenvalue is $0$, and then rescaling such that the largest eigenvalue is $1$. We refer to this ``normalized'' version of GUE as nGUE. It is easy to see that, as above, when $\widehat{H}$ is sampled from nGUE, $\tr \widehat{H}$ concentrates for $d \gg 1$. Moreover, since $\av{G + G^\dagger}_{\mathrm{GUE}} = 0$,
\begin{align*}
    \frac{1}{d} \av{\tr \widehat{H}}_{\mathrm{nGUE}} = \frac{1}{2}.
\end{align*}
This means that the condition in Eq.~\eqref{Ham_cond}, with $\hat{\mathfrak{h}} = 1/2$, is satisfied for almost all $\widehat{H}$'s sampled from nGUE. Note that the nGUE ensemble contains not only $r$-local $\widehat{H}$'s, but also highly nonlocal ones. This however does not qualitatively affect the concentration properties as they are determined by the spectral spread of the Hamiltonian [Eqs.~\eqref{conc_erg} and~\eqref{width_erg}] and $\spr(\widehat{H}) = 1$ by definition. The $r$-dependence enters into the picture via $H = \spr(H) \widehat{H}$, and for an $r$-local many-body system, $\spr(H) \propto N^r$ [Eq.~\eqref{ham_scal_gen}].

\begin{figure}[!t]
    \centering
    \includegraphics[trim=0.25cm 0.2cm 0.1cm 0.2cm, clip, width=\columnwidth]{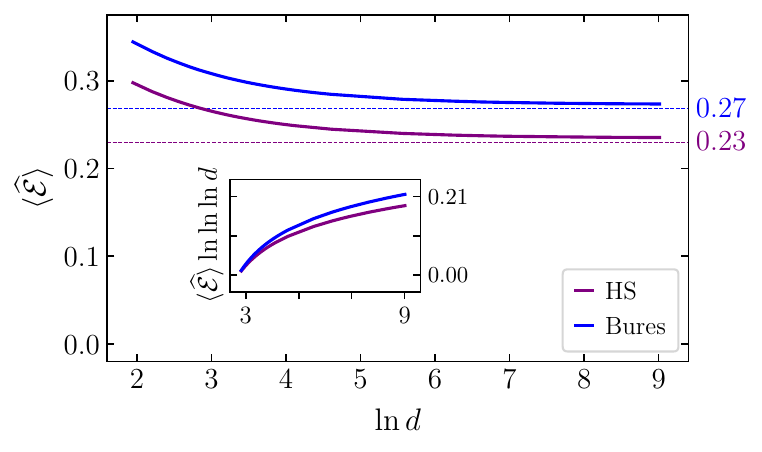}
    \caption{The average ergotropy $\avHS{\widehat{\erg}}$ [Eq.~\eqref{Erg_hat}] as a function of the Hilbert-space dimension $d$, with $\rho$ sampled from the Hilbert--Schmidt and Bures measures. The normalized Hamiltonian $\widehat{H}$ is drawn from nGUE, and $d$ ranges from $2^3$ to $8200 \gtrsim 2^{13}$. For $d \leq 260$, each point on the plot is computed on an ensemble of size $10^6$. For $d \geq 520$, the size of each ensemble is only $150$, which, due to the exponential concentration, is large enough to reliably retrieve the averages.
    The plot shows that both $\avHS{\widehat{\erg}}$ and $\avBu{\widehat{\erg}}$ tend to constant nonzero values for large values of $d$.
    To demonstrate that $\av{\widehat{\erg}}$ does not keep decaying with $d$, the \textbf{inset} shows $\av{\widehat{\erg}} \ln \ln \ln d$ as a function of $\ln d$. Since the growth of $\ln \ln \ln d$ is extremely slow, the monotonic increase of $\av{\widehat{\erg}} \ln \ln \ln d$ strongly supports the hypothesis that $\av{\widehat{\erg}}$ indeed tends to a constant in the $d \to \infty$ limit.
    The dashed lines show the corresponding projected asymptotic values.
    }
    \label{fig:av_erg}
\end{figure}

The results of our numerical analysis are presented in Fig.~\ref{fig:av_erg} for both Hilbert--Schmidt and Bures (see Sec.~\ref{sec:Bures}) measures. As can be seen there,
\begin{align} \label{av_erghat}
     \avHS{\widehat{\erg}} \gtrsim 0.23 \qquad \mathrm{and} \qquad \avBu{\widehat{\erg}} \gtrsim 0.27,
\end{align}
which, via Eq.~\eqref{hat_to_original}, establishes that
\begin{align} \label{av_true_erg}
\begin{split}
    \avHS{\erg\big(\rho, H \big)} \gtrsim 0.23 \, \spr(H) ,
    \\[3pt]
    \avBu{\erg\big(\rho, H \big)} \gtrsim 0.27 \, \spr(H) .
\end{split}
\end{align}
Note that, even if $\avHS{\widehat{\erg}}$ and $\avBu{\widehat{\erg}}$ in Fig.~\ref{fig:av_erg} were to keep decaying and eventually go to zero, they would have to do so more slowly than $\frac{1}{\ln \ln N}$. In which case, they can still be considered constant for all practical purposes since $\ln \ln N$ grows extremely slowly. E.g., for the Avogadro number $N_{\mathrm{A}} \approx 6.022 \times 10^{23}$, $\ln \ln N_{\mathrm{A}}\approx 4$, whereas $e^{e^5}$ is already $\gg N_{\mathrm{A}}^{2.5}$ and $e^{e^6}$ is much larger than the estimated number of particles in the universe.

For generic many-body systems, $\spr(H)$ is macroscopic, at most scaling as $N^r$ [see Eq.~\eqref{ham_scal_gen}]. Thus, Eqs.~\eqref{av_true_erg} and~\eqref{avent_scaling} show that the system has \textit{macroscopic ergotropy and entropy at the same time}. If the states were sampled from an ensemble of pure states (e.g., the Fubini--Study measure), then the macroscopic average of the ergotropy would not be surprising at all---the ergotropy of pure states simply coincides with their average energy. Here, however, the states are general and typically have a full rank and macroscopic entropy, and there is no a priori expectation that these states could store significant amounts of ergotropy. Therefore, it is surprising to find that these highly mixed states do indeed come with a macroscopic ergotropy.

\section{Concentration under the Bures measure}
\label{sec:Bures}

The coincidence that, out of all the ``major'' distance functions, entropy and ergotropy happen to be simultaneously Lipschitz continuous only with respect to the Bures distance adds another entry to the catalog of its unique properties. Namely, the Bures distance is the only metric that is simultaneously Riemannian, Fubini--Study--adjusted, contractive, and Fisher-adjusted \cite{Petz_1996a, Sommers_2003, Bengtsson_book_2006}. This is already a good motivation to study the concentration properties of $\erg(\rho, H)$ and $S(\rho)$ when $\rho$ is drawn from the Bures measure---the normalized Riemannian volume measure induced by the Bures distance \cite{Bengtsson_book_2006}.

\begin{figure*}[!t]
    \centering
    \includegraphics[trim=0.31cm 0.35cm 0.23cm 0.1cm, clip, width=\textwidth]{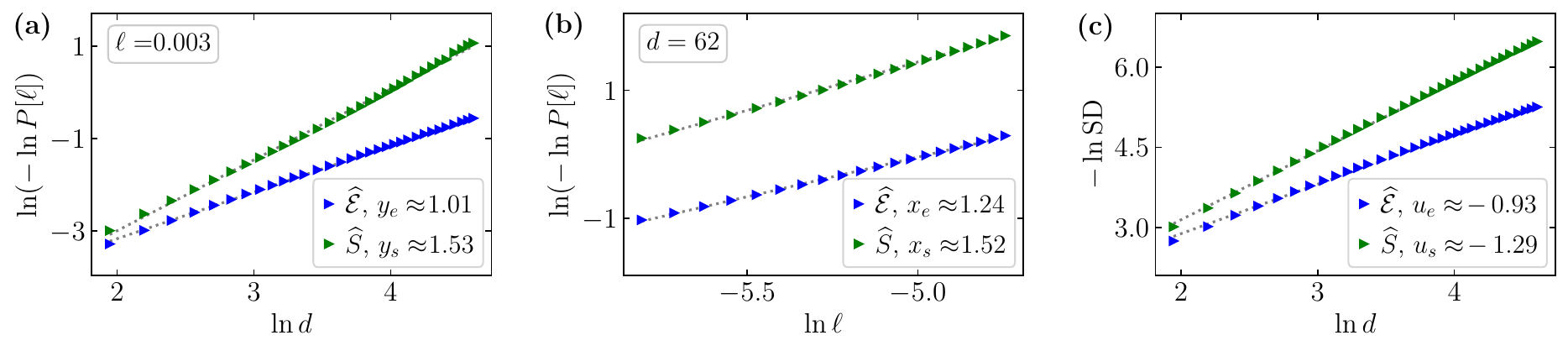}
    \caption{The concentration properties of ergotropy and entropy when the state is drawn from the Bures measure.
    Panel \textbf{(a)} shows the dependence of the deviation probabilities $P_{\widehat{\erg}}[\ell]$ and $P_{\widehat{S}}[\ell]$ on $d$ when $\ell$ is fixed [Eq.~\eqref{numform}]. Panel \textbf{(b)} shows Eq.~\eqref{numform} when $d$ is fixed and $\ell$ is varied. Panel \textbf{(c)} shows the dependence on the standard deviations (SD) of $\widehat{\erg}$ and $\widehat{S}$ as functions of $d$ [Eq.~\eqref{SD_scaling}]. The slope of each curve gives its respective exponent ($y_e$, $y_s$, $x_e$, $x_s$, $u_e$, or $u_s$). As in Fig.~\ref{fig:av_erg}, the Hamiltonian $\widehat{H}$ is sampled from nGUE. Each point on on each line is calculated on an ensemble of size $10^6$. The dotted gray lines are linear fits, from which the slopes are calculated.
    }
    \label{fig:Bures_concentration}
\end{figure*}

A deeper reason to study this problem comes from the distinguished inference-theoretical status of the Bures measure, which is often regarded as the least informative prior over quantum states. While this notion is challenging to pin down conceptually \cite{Jaynes_1968, Amari-Nagaoka_book_2000}, there is a fairly definitive answer in classical statistics. There, the Jeffreys prior satisfies a number of desirable conditions \cite{Jaynes_1968} and is widely accepted to be the least informative prior invariant under reparametrizations of the classical Fisher metric \cite{Jaynes_1968, Amari-Nagaoka_book_2000}. On the manifold of quantum states, the analog of the Jeffreys prior is the Bures measure \cite{Slater_1996, Hall_1998, Slater_1998b, Sommers_2003, Bengtsson_book_2006} (for a brief explanation why, see Appendix~\ref{app:bures}).

Despite these special properties of the Bures measure, to the best of our knowledge, no rigorous measure-concentration bounds have yet been established for it. The standard geometric approach to deriving such bounds relies on a global lower bound on the Ricci curvature of the underlying Riemannian metric (here, the Bures metric on the manifold of density matrices) \cite{Milman_book_2001, Ledoux_book_2001}. However, no suitable Ricci-curvature bound is currently known for the Bures metric \cite{Dittmann_1995, Dittmann_1999, Slater_2002, Slater_2003, Slater_2005, Bengtsson_book_2006}. Nonetheless, the Bures measure is amenable to efficient numerical simulations. Indeed, as proven in Ref.~\cite{Osipov_2010}, the states
\begin{align} \label{bures_rs}
    \rho = \frac{(\id + U) G G^\dagger (\id + U^\dagger)}{\tr \big[ (\id + U) G G^\dagger (\id + U^\dagger) \big]},
\end{align}
where $G$ is a $d\times d$ random matrix drawn from the Ginibre ensemble and $U$ is a random unitary matrix drawn from the Haar measure on the $U(d)$ group \cite{Mezzadri_2007}, are distributed according to the Bures measure.


Our extensive numerical analysis, based on Eq.~\eqref{bures_rs} and presented in Fig.~\ref{fig:Bures_concentration}, demonstrates that both $\erg(\rho, \widehat{H})$ and $S(\rho)$ concentrate around their ensemble averages, $\avBu{\erg(\rho, \widehat{H})}$ and $\avBu{S(\rho)}$, when $\rho$ is sampled from the Bures measure and $\widehat{H}$ from nGUE. Moreover, as shown in Fig.~\ref{fig:av_erg}, $\avBu{\erg(\rho, \widehat{H})}$ approaches a finite, nonzero value [Eq.~\eqref{av_erghat}] as $d \to \infty$.

To directly probe this concentration phenomenon, we define the deviation probabilities
\begin{align}
\begin{split}
    P_{\widehat{\erg}}[\ell] &\coloneq \prob\Big[\big\vert \erg(\rho, \widehat{H}) - \avBu{\widehat{\erg}} \big\vert > \ell \Big],
    \\
    P_{\widehat{S}}[\ell] &\coloneq \prob\Big[\big\vert \widehat{S}(\rho) - \avBu{\widehat{S}} \big\vert > \ell \Big].
\end{split}
\end{align}
Here, $\widehat{S}(\rho) \coloneq S(\rho) / \ln d$ is introduced so that $\widehat{S}$ and $\widehat{\erg}$ are on the same footing in the sense that both $L^{\mathrm{B}}_{\widehat{\erg}}$ and $L^{\mathrm{B}}_{\widehat{S}}$ are $O(1)$ [see Eqs.~\eqref{erg_lipschs} and~\eqref{entropy_lips_Bu_coeff}].

Now, motivated by the Lévy-type bounds in Eqs.~\eqref{conc_erg} and~\eqref{conc_ent}, we hypothesize that
\begin{align} \label{hyphoB}
\begin{split}
    P_{\widehat{\erg}}[\ell] &\lesssim \xi_e \exp\big(\!- \theta_e \, \ell^{x_e} d^{y_e}\big),
    \\
    P_{\widehat{S}}[\ell] &\lesssim \xi_s \exp\big(\!- \theta_s \ell^{x_s} d^{y_s}\big),
\end{split}
\end{align}
where $\xi$'s and $\theta$'s are positive constants, and $x_e$, $y_e$, $x_s$, and $y_s$ are positive exponents. This ansatz implies
\begin{align} \label{numform}
    \ln(-\ln P[\ell]) \approx \mathrm{const} + x \ln \ell + y \ln d.
\end{align}
For both $\widehat{\erg}$ and $\widehat{S}$, this relation is well supported by the numerical data shown in Fig.~\ref{fig:Bures_concentration}(a,b). The observed linear behavior provides strong evidence for the validity of Eq.~\eqref{hyphoB} and allows us to extract numerical estimates of the exponents $x_e$, $y_e$, $x_s$, and $y_s$. Note, however, that Eq.~\eqref{numform} effectively assumes $\xi_e \approx 1$ and $\xi_s \approx 1$, which is likely not the case. The slight deviations from linearity visible in Fig.~\ref{fig:Bures_concentration}(a,b) may therefore be attributable, at least in part, to this approximation.

We next examine how the standard deviations (SD) of $\widehat{\erg}$ and $\widehat{S}$ depend on the dimension $d$. Equation~\eqref{hyphoB} implies that SDs should decay polynomially with $d$, corresponding to the scaling form
\begin{align} \label{SD_scaling}
    - \ln \mathrm{SD}_{\widehat{\erg}, \widehat{S}} = \mathrm{const} + u_{e, s} \ln d,
\end{align}
where $u_e$ and $u_s$ are positive exponents. Figure~\ref{fig:Bures_concentration}(c) exhibits precisely this linear scaling behavior, providing further evidence for concentration of both quantities as the dimension increases.

We have thus shown that both ergotropy and entropy exhibit exponential concentration around their ensemble averages also under the Bures measure. Remarkably, however, our numerical results indicate that this concentration is \textit{not} governed by a Lévy-like bound [Eq.~\eqref{Levis}]. Indeed, the estimates obtained for the exponents satisfy $x_e, \, y_e, \,x_s, \,y_s \, < 2$, implying a slower decay of the corresponding deviation probabilities than the Lévy scaling $e^{-\mathrm{const} \, \ell^2 d^2}$. Thus, while concentration persists, its scaling appears to be different from that predicted by Lévy's lemma. More broadly, these findings suggest that the concentration properties of the Bures measure may indeed not be governed by a Lévy-type mechanism, pointing instead to a distinct concentration structure that remains to be understood.

\section{Fluctuations within ergotropy}

\begin{figure*}[t!]
    \centering
    \includegraphics[width=0.99\linewidth]{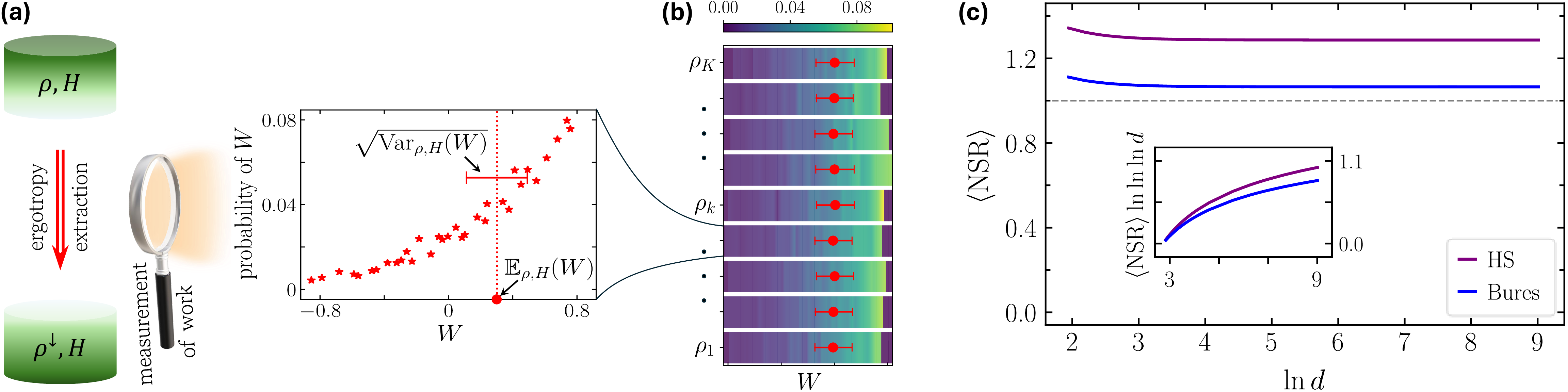}
    \caption{
    Intrinsic versus ensemble fluctuations of ergotropy. Panel \textbf{(a)} shows the measurement statistics of work when the ergotropy is extracted from a system with $d = 33$, initially prepared in a randomly generated state $\rho$. Work is measured using the work operator scheme, resulting in a broad probability distribution. Panel \textbf{(b)} shows the corresponding work distributions, now displayed as one-dimensional heatmaps, for several states $\rho_k$ sampled from the Hilbert--Schmidt measure; the Hamiltonian $H$ is the same for all of them. The red circles mark the averages $\mathbbm{E}_{\rho_k, H}(W) = \erg(\rho_k, H)$, and the bars denote the standard deviations $\sqrt{\var_{\rho_k, H}(W)}$. Despite the substantial width of each individual distribution, their means and standard deviations vary only weakly from state to state. This illustrates the concentration of measure: while work exhibits significant fluctuations for each $\rho$, the statistical properties of these fluctuations are nearly independent of $\rho$.
    Panel \textbf{(c)} shows the ensemble average of NSR---the noise-to-signal ratio defined in Eq.~\eqref{defNSR}---as a function of $d$. The same data and analysis as in Fig.~\ref{fig:av_erg} are used here. We see that $\av{\mathrm{NSR}}$ is slightly above $1$ (dashed line) for both Hilbert--Schmidt and Bures measures, indicating strong intrinsic fluctuations of work in all dimensions. As in Fig.~\ref{fig:av_erg}, the inset demonstrates that $\av{\mathrm{NSR}}$ remains finite in the $d \to \infty$ limit.
    }
    \label{fig:quantum_vs_ensemble}
\end{figure*}

So far, we have focused on the ergotropy $\erg (\rho, H)$ and its statistics when $\rho$ is drawn at random. However, $\erg (\rho, H)$ is an expectation value in itself---for each fixed $\rho$, it is the \textit{average} work extracted by evolving the system by the unitary $U_{\rho^\downarrow}$ [Eq.~\eqref{ergot_def_1}]. To make this operationally meaningful, one needs to specify how extracted work would be observed. Any faithful work measurement scheme would then yield a random outcome $W$ with probability distribution $P_{\rho, H}(W)$, whose expectation value reproduces the ergotropy:
\begin{align} \label{faithfulness}
    \mathbbm{E}_{\rho, H}[W] \coloneqq \sum_W P_{\rho, H}(W) W = \erg(\rho, H).
\end{align}
The fluctuations of work around this average can be characterized by the variance $\var_{\rho, H}(W) = \mathbbm{E}_{\rho, H}[W^2] - \mathbbm{E}_{\rho, H}[W]^2$, which generally depends on the details of the measurement scheme. However, faithfully measuring work for initially coherent systems is a well-known challenge as there is no unique, well-defined way of doing that \cite{Allahverdyan_2005, Talkner_2016, Miller_2017, Baumer_2018, Pei_2023, Lostaglio_2023, Hovhannisyan_2024a}. In particular, the standard two-point measurement (TPM) scheme \cite{Kurchan_2000, Tasaki_2000} cannot be used to characterize the fluctuations of work when it is extracted from states with energetic coherences \cite{Allahverdyan_2005, Talkner_2016, Kammerlander_2016, Perarnau_2017, Hovhannisyan_2024a}. This problem is relevant here because, as noted above, almost all $\rho$'s drawn from the Hilbert--Schmidt and Bures measures are coherent in the energy eigenbasis.

Luckily, ergotropy is somewhat unique in that, although the initial state may have coherences, the final passive state is always diagonal in the energy eigenbasis. Due to this, the reverse TPM scheme introduced in Ref.~\cite{Hovhannisyan_2024a} and the so-called dynamic Bayesian network scheme \cite{Park_2017, Micadei_2020} (both state-dependent schemes), as well as the Margenau--Hill--Terletskii distribution introduced in Ref.~\cite{Allahverdyan_2014} (a quasiprobability distribution from the Kirkwood--Dirac class \cite{Pei_2023, Lostaglio_2023}), produce the same statistics for work. Moreover, the first two moments of all these statistics coincide with those of the Heisenberg operator of work scheme \cite{Bochkov_1977, Allahverdyan_2005} (a state-independent scheme). All these schemes reproduce the ergotropy as the average extracted work and yield the same variance
\begin{align}
    \var_{\rho, H}(W) = \tr\big[\rho \big(H - U_{\rho^\downarrow}^\dagger H U_{\rho^\downarrow}^{\phantom{\dagger}}\big)^2\big] - \erg(\rho, H)^2.
\end{align}
We adopt this common expression as the work variance for ergotropy extraction. Furthermore, to capture the strength of these fluctuations, we introduce the ``noise-to-signal ratio'' (NSR),
\begin{align} \label{defNSR}
    \mathrm{NSR}_{\rho, H} \coloneq \frac{\sqrt{\var_{\rho, H}(W)}}{\mathbbm{E}_{\rho, H}[W]}.
\end{align}
This quantity is always well-defined as $\var_{\rho, H}(W) > 0$ necessitates $\mathbbm{E}_{\rho, H}[W] = \erg(\rho, H) > 0$. Conveniently, NSR is also insensitive to the norm of $H$ in the sense that $\mathrm{NSR}_{\rho, H} = \mathrm{NSR}_{\rho, \widehat{H}}$.

We emphasize that the ``intrinsic'' fluctuations of work, characterized by $\var_{\rho, H}(W)$, are not to be confused with the ``ensemble'' fluctuations of $\erg(\rho, H)$ itself when $\rho$ is randomized. This distinction is illustrated in Fig.~\ref{fig:quantum_vs_ensemble}(a,b). There, we see that, while the first two moments of work, $\erg(\rho, H)$ and $\var_{\rho, H}(W)$, are nearly independent of the state due to suppressed ensemble fluctuations [Eqs.~\eqref{conc_erg} and~\eqref{conc_erg_width}], $\var_{\rho, H}(W)$ is comparable to $\erg(\rho, H)$ indicating substantial intrinsic fluctuations. Our numerical simulations further show that, like ergotropy, both $\mathrm{NSR}_{\rho, H}$ and $\var_{\rho, H}(W)$ concentrate in high Hilbert-space dimensions under the Hilbert--Schmidt and Bures measures. Fig.~\ref{fig:quantum_vs_ensemble}(c) presents the ensemble average of $\mathrm{NSR}_{\rho, \widehat{H}}$, simply denoted by $\av{\mathrm{NSR}}$, for different dimensions. For both Hilbert--Schmidt and Bures measures, we find that $\av{\mathrm{NSR}} \gtrsim 1$ across all dimensions. Thus, typically, the intrinsic fluctuations of ergotropy are rather strong even at high dimensions, where ensemble fluctuations are practically nonexistent.

\section{Discussion and outlook}

We have shown that most states of a many-body battery simultaneously accommodate a macroscopic amount of ergotropy [Eq.~\eqref{av_true_erg}] and entropy [Eq.~\eqref{avent_scaling}]. This finding suggests that batteries made of a very broad class of many-body systems are substantially more robust than one would naively anticipate. Indeed, our results imply that maintaining a large charge does not require extreme fine-tuning of the charge-holding state. From a practical viewpoint, an important consequence is that once the system is prepared in a state with close-to-average ergotropy, small perturbations are likely to keep it within the typical subspace. This can be interpreted as the stored charge exhibiting robustness against noise, as sampling states from a measure models a noisy source preparing a different state in each realization. Of course, the ergotropy-extracting unitary generally depends on the state itself, and understanding the sensitivity of the optimal extraction protocol to small state perturbations is an interesting direction for future work.

It should be noted that our results do not necessarily mean that essentially any physical object around us harbors a macroscopic amount of ergotropy. The states of typical many-body systems encountered in nature are unlikely to be distributed according to either the Hilbert--Schmidt or Bures measure. In particular, realistic many-body systems tend to thermalize, and such states of realistic Hamiltonians occupy only a small corner of the Hilbert space (see, e.g., Ref.~\cite{Cirac_2009}).

That said, thermalization generally concerns local subsystems rather than the global many-body state \cite{Gemmer_2004, Gogolin_2016, Mori_2018, Hovhannisyan_2023, Bertoni_2025}. The global state may thus retain a substantial degree of effective randomness. For moderately sized large (mesoscopic) systems, its ``full'' randomization is not therefore implausible, and a suitable environment may render it random enough that macroscopic ergotropy becomes typical. By ``random enough,'' we mean that macroscopic ergotropy need not be typical only for the Hilbert--Schmidt and Bures measures: its occurrence for both suggests that the same behavior may persist for a broader class of ensembles. Identifying the degree and form of randomness required for macroscopic typical ergotropy, as well as engineering battery--environment interactions capable of generating it, are important problems in their own right.

In this context, we note that the problem considered here is naturally connected to work extraction from partially or fully unknown states \cite{Safranek_2023, Watanabe_2024, Chakraborty_2025}. Exploring the implications of our results for that setting represents another interesting direction for future research.

More generally, the concentration of measure phenomenon has already found many profound applications in quantum information theory (e.g., the so-called ``probabilistic method'') \cite{Hayden_2006, Watrous_book_2018} and quantum statistical mechanics \cite{Popescu_2006, Gogolin_2016}. We therefore believe that adding ergotropy to the list of meaningful Lipschitz continuous quantities will extend the scope of these methods to the quantum thermodynamics of many-body systems, even beyond quantum batteries. Finally, our numerics indicate exponential concentration of entropy and ergotropy under the Bures measure through a mechanism distinct from that underlying Lévy’s lemma. This raises the intriguing possibility of a deeper mathematical structure governing concentration.

\section*{Acknowledgments}

We thank Zhirayr Avetisyan for illuminating discussions on measure concentration.
K.H. and J.A. are grateful for support from the University of Potsdam.
J.A. gratefully acknowledges funding from the Deutsche Forschungsgemeinschaft (DFG, German Research Foundation) under Grants No. 384846402 and No. 513075417 and from the Engineering and Physical Sciences Research Council (EPSRC) (Grant No. EP/R045577/1) and thanks the Royal Society for support.

\medskip

The code used to produce the data shown in the figures is available upon reasonable request to K.H., \href{mailto:karen.hovhannisyan@uni-potsdam.de}{karen.hovhannisyan@uni-potsdam.de}.

For the purpose of open access, the authors have applied a `Creative Commons Attribution' (CC BY) license to any Author Accepted Manuscript version arising from this submission.

\bigskip


\appendix


\section{Lipschitz-continuity of ergotropy}
\label{app:erg_cont}

Here we prove Theorem~\ref{thm:ergotropy_lipschitz} in the main text. To show that $\erg(\rho)$ is Lipschitz continuous, we invoke Eq.~\eqref{ergot_def_2} and write
\begin{align} \nonumber
    \vert \erg(\rho, H) - \erg(\sigma, H) \vert &= \big\vert \! \tr[(\rho - \sigma) H] - \tr[(\rho^\downarrow - \sigma^\downarrow) H] \big\vert
    \\ \label{erg_lip_1}
    &\leq \big\vert \!\tr[(\rho - \sigma) H] \big\vert + \big\vert \! \tr[(\rho^\downarrow - \sigma^\downarrow) H] \big\vert . ~
\end{align}
In what follows, we will upper-bound both terms in Eq.~\eqref{erg_lip_1} in terms of the distance between $\rho$ and $\sigma$.

Recall that, by our convention, the eigenvalues $E_i$ of $H$ are always listed increasingly, so $\spr(H) = E_{d-1} - E_0$. Furthermore, since $\tr(\rho - \sigma) = 0$, we can equivalently write Eq.~\eqref{erg_lip_1} with the constant-shifted Hamiltonian
\begin{align} \label{erg_lip_3}
    H' = H - \frac{\spr(H)}{2} \id
\end{align}
instead of $H$. Obviously,
\begin{align} \label{erg_lip_4}
    \opn{H'} = \frac{1}{2}\spr(H).
\end{align}
Now, let $\lambda(A)_i$ denote the eigenvalues of the operator $A$. Then, in the eigenbasis of $\rho - \sigma$, we have
\begin{align} \nonumber
    \vert \tr[(\rho - \sigma) H]\vert &= \big\vert \tr[(\rho - \sigma) H'] \big\vert
    \\ \nonumber
    &= \Big\vert \sum_i \lambda(\rho - \sigma)_i H'_{ii} \Big\vert
    \\ \nonumber
    &\leq \sum_i \vert \lambda(\rho - \sigma)_i \vert \cdot \vert H'_{ii} \vert
    \\ \label{erg_lip_5}
    &\leq \opn{H'} \sum_i \vert \lambda(\rho - \sigma)_i \vert
    \\ \label{erg_lip_6}
    &= 2 \opn{H'} \, \DTr(\rho, \sigma).
\end{align}
Inequality~\eqref{erg_lip_5} follows from the fact that, by definition, $\vert H'_{ii} \vert \leq \opn{H'}$, $\forall i$. When transitioning from Eq.~\eqref{erg_lip_5} to Eq.~\eqref{erg_lip_6}, we used the fact that the definition of the trace distance in Eq.~\eqref{DTr_def} is equivalent to
\begin{align} \label{erg_lip_7}
    \DTr(\rho, \sigma) = \frac{1}{2} \sum_i \vert \lambda(\rho - \sigma)_i \vert.
\end{align}
Using the Fuchs--van de Graaf inequalities \cite{Fuchs_1999}, it is straightforward to show that
\begin{align} \label{HoFuGra}
    \frac{1}{2} \DBu(\rho, \sigma)^2 \leq \DTr(\rho, \sigma) \leq \DBu(\rho, \sigma).
\end{align}
Moreover, a standard relation between Schatten norms entails (see, e.g., Ref.~\cite{Coles_2019})
\begin{align} \label{HS_TrD}
    \DTr(\rho, \sigma) \leq \sqrt{d/4} \, \DHS(\rho, \sigma).
\end{align}
We note in passing that the lower bound in Eq.~\eqref{HoFuGra} was first proven in Ref.~\cite{Bures_1969}; see also \cite{Kholevo_1972, Wilde_2018}.

Summarizing the analysis between Eqs.~\eqref{erg_lip_4} and~\eqref{HS_TrD}, we have the following upper bounds for the first term in Eq.~\eqref{erg_lip_1}:
\begin{align} \label{lips_1}
\vert \tr[(\rho - \sigma) H] \vert \leq \left\{ \begin{array}{l} \vspace{2mm}
      \spr(H) \, \DTr(\rho, \sigma),
    \\ \vspace{2mm}
    \spr(H) \, \DBu(\rho, \sigma),
    \\
    \sqrt{d/4} \, \spr(H) \, \DHS(\rho, \sigma). \end{array} \right.
\end{align}

Turning to the second term in Eq.~\eqref{erg_lip_1}, let us write it in the energy eigenbasis:
\begin{align} \label{erg_lip_9}
\begin{split}
    \big\vert \tr[(\rho^\downarrow - \sigma^\downarrow) H] \big\vert &= \big\vert \tr[(\rho^\downarrow - \sigma^\downarrow) H'] \big\vert
    \\
    &= \Big\vert \sum_i \big[\lambda(\rho)^\downarrow_i - \lambda(\sigma)^\downarrow_i\big] \, E'_i \Big\vert
    \\
    &\leq \sum_i \big\vert \lambda(\rho)^\downarrow_i - \lambda(\sigma)^\downarrow_i \big\vert \cdot \big\vert E'_i \big\vert
    \\
    &\leq \opn{H'} \sum_i \big\vert \lambda(\rho)^\downarrow_i - \lambda(\sigma)^\downarrow_i \big\vert.
\end{split}
\end{align}
To proceed further, we invoke the Lidskii--Wielandt theorem \cite{Lidskii_1950} in its representation as Eq.~(IV.62) in Ref.~\cite{Bhatia_book_1997} (see also Ref.~\cite{Mirsky_1960}).
For the trace norm, this theorem states that
\begin{align} \label{erg_lip_10}
    \frac{1}{2} \sum_i \big\vert \lambda(\rho)^\downarrow_i - \lambda(\sigma)^\downarrow_i \big\vert \leq \DTr(\rho, \sigma).
\end{align}
Plugging Eq.~\eqref{erg_lip_10} into Eq.~\eqref{erg_lip_9}, and keeping in mind Eq.~\eqref{HoFuGra}, we obtain
\begin{align} \label{lips_2}
\begin{split}
    \big\vert \tr[(\rho^\downarrow - \sigma^\downarrow) H] \big\vert &\leq \spr(H) \, \DTr(\rho, \sigma)
    \\
    &\leq \spr(H) \, \DBu(\rho, \sigma).
\end{split}
\end{align}
Also, taking into account Eq.~\eqref{HS_TrD}, we have
\begin{align} \label{lips_3}
\begin{split}
    \big\vert \tr[(\rho^\downarrow - \sigma^\downarrow) H] \big\vert &\leq \opn{H} \, \DTr(\rho, \sigma)
    \\
    &\leq \sqrt{d/4} \, \spr(H) \, \DHS(\rho, \sigma).
\end{split}
\end{align}

\medskip

Finally, together with Eqs.~\eqref{lips_1},~\eqref{lips_2}, and~\eqref{lips_3}, Eq.~\eqref{erg_lip_1} leads us to
\begin{align} \label{erg_lips}
    \vert \erg(\rho, H) - \erg(\sigma, H)| \leq \left\{ \begin{array}{l} \vspace{2mm}
    2 \, \spr(H) \, \DTr(\rho, \sigma),
    \\ \vspace{2mm}
    2 \, \spr(H) \, \DBu(\rho, \sigma),
    \\
    \sqrt{d} \, \spr(H) \, \DHS(\rho, \sigma). \end{array} \right.
\end{align}
These inequalities thus prove that $\erg(\rho, H)$, as a function of $\rho$, is Lipschitz continuous with respect to the trace, Bures, and Hilbert--Schmidt distances. By definition, the inequalities in Eq.~\eqref{erg_lips} also provide upper bounds for the respective Lipschitz constants $L_{\mathrm{Tr}}^\erg$, $L_{\mathrm{B}}^\erg$, and $L_{\mathrm{HS}}^\erg$:
\begin{align}
\begin{array}{rl} \vspace{2mm}
    L_{\mathrm{Tr}}^\erg \!\!\! & \leq 2 \, \spr(H),
    \\ \vspace{2mm}
    L_{\mathrm{B}}^\erg \!\!\! & \leq 2 \, \spr(H),
    \\
    L_{\mathrm{HS}}^\erg \!\!\! & \leq \sqrt{d} \, \spr(H).
\end{array}
\end{align}
This thus concludes the proof of Theorem~\ref{thm:ergotropy_lipschitz}.

\medskip

Lastly, let us prove Corollary~\ref{thm:ergotropy_lipschitz_purified}, which extends the continuity bounds above to the purification space consisting of states $\ket{\P(\rho)}$. To do so, we note that, due to Uhlmann's theorem \cite{Uhlmann_1976},
\begin{align} \label{UhlmannT}
    F(\rho, \sigma) \geq \big\vert \! \braket{\P(\rho)}{\P(\sigma)} \! \big\vert^2.
\end{align}
Therefore, in view of Eq.~\eqref{DBu_def},
\begin{align} \label{bures-leq-euclid}
\begin{split}
    \DBu(\rho, \sigma) &\leq \sqrt{2 - 2 \, \vert \! \braket{\P(\rho)}{\P(\sigma)} \! \vert}
    \\
    &\leq \sqrt{2 - 2 \, \mathrm{Re} \braket{\P(\rho)}{\P(\sigma)}}
    \\
    &= \Vert \ket{\P(\rho)} - \ket{\P(\sigma)} \! \Vert_2 \, ,
\end{split}
\end{align}
where $\Vert \ket{\phi} \Vert_2 = \sqrt{\braket{\phi}{\phi}}$ is the Schatten $2$-norm, which, for vectors, is better known as the Euclidean norm. Plugging Eq.~\eqref{bures-leq-euclid} into the second line of Eq.~\eqref{erg_lips} yields Eq.~\eqref{koxac1}, thereby proving Corollary~\ref{thm:ergotropy_lipschitz_purified}.

\section{Lipschitz-continuity of entropy}
\label{app:ent_cont}

Let us show how to obtain the bound in Eq.~\eqref{entropy_lips_Bu} from the bound proven in Ref.~\cite{Sekatski_2021}. There, it was established that
\begin{align} \label{pablo}
|S(\rho) - S(\sigma)| \leq L^{\mathrm{BA}}_S \, \arccos \sqrt{F(\rho, \sigma)},
\end{align}
with
\begin{align} \label{Cd_1}
L^{\mathrm{BA}}_S \leq \left\{
\begin{array}{ll} \vspace{2mm}
\frac{1.609}{\ln 2} \sqrt{d-1}, & \;\mathrm{for} \;\; d \leq 4
\\
\frac{2}{\ln 2} \ln d, & \;\mathrm{for} \;\; d \geq 5
\end{array}
\right. .
\end{align}
Since here we are mainly concerned with high-dimensional systems, we will need only the second row of Eq.~\eqref{Cd_1}. (Note that, as can be easily checked numerically, $L^{\mathrm{BA}}_S < 3.35 \, \ln d$, $\forall d \geq 2$.)

Now, from the identity $\cos x = 1 - 2 \sin^2 (x/2)$ it straightforwardly follows that
\begin{align*}
    \arccos (1 - 2 x^2) = 2 \arcsin x.
\end{align*}
Furthermore, due to $\sin x \geq 2 x / \pi$ ($x \in [0, \pi/2]$), the inequality $\arcsin x \leq \pi x / 2$ holds. Using these and the definition of the Bures distance given in Eq.~\eqref{DBu_def}, we can write
\begin{align} \label{BuA_BuD}
\begin{split}
\arccos \sqrt{F(\rho, \sigma)} &= \arccos \big(1- \DBu(\rho, \sigma)^2 / 2\big)
\\
&= 2 \arcsin \frac{\DBu(\rho, \sigma)}{2}
\\
&\leq \frac{\pi}{2} \DBu(\rho, \sigma).
\end{split}
\end{align}
Plugging this into Eq.~\eqref{pablo}, and assuming $d \geq 5$, we thus obtain
\begin{align} \label{entropy_lips_Bu_app}
|S(\rho) - S(\sigma)| \leq \frac{\pi \ln d}{\ln 2} \, \DBu(\rho, \sigma),
\end{align}
which coincides with Eq.~\eqref{entropy_lips_Bu}.

\medskip

Let us note in passing that, using the lower bound in the Fuchs--van de Graaf inequalities~\eqref{HoFuGra} and the relation between Schatten norms in Eq.~\eqref{HS_TrD}, one can obtain from Eq.~\eqref{entropy_lips_Bu} that
\begin{align} \label{entropy_lips_TrSH}
\begin{split}
|S(\rho) - S(\sigma)| &\leq \sqrt{2} \, L_{\mathrm{B}}^S \, \DTr(\rho, \sigma)^{1/2}
\\
&\leq d^{1/4} \, L_{\mathrm{B}}^S \, \DHS(\rho, \sigma)^{1/2}.
\end{split}
\end{align}
This means that the von Neumann entropy is H\"{o}lder continuous with exponent $1/2$ with respect to the trace and Hilbert--Schmidt distances.

Equations~\eqref{entropy_lips_Bu} and~\eqref{entropy_lips_TrSH} give an unusual perspective on the continuity of entropy compared to the ``classic'' bounds such as the one in Eq.~\eqref{entropy_cont} (see also Ref.~\cite{Winter_2016}).

\section{Applying Levy's lemma to ergotropy and entropy}
\label{app:hilbert-to-sphere}

Here we show how concentration of the Haar measure on the sphere translates into concentration of the Hilbert--Schmidt measure on density matrices.

As mentioned in the main text, if $\ket{\psi}$ is sampled from the Fubini--Study measure on $\mathcal{H}_d \otimes \mathcal{H}_d$, then $\T(\ket{\psi})$ is distributed according to the Hilbert--Schmidt measure on $\mathcal{H}_d$ \cite{Zyczkowski_2001}. Furthermore, there is a natural one-on-one correspondence between the Fubini--Study measure on pure states in $\mathcal{H}_d \otimes \mathcal{H}_d$ and the Haar measure on $\mathbbm{S}^{2 d^2 - 1}$ \cite{Bengtsson_book_2006} that can be constructed as follows. Fix an arbitrary basis $\{\ket{\zeta}\}_{\zeta = 0}^{d^2 - 1}$ in $\mathcal{H}_d \otimes \mathcal{H}_d$, such that any pure state $\ket{\phi}$ in it decomposes into $\ket{\phi} = \sum_{\zeta} \bm{\phi}_\zeta \ket{\zeta}$. Then, the normalization $\braket{\phi}{\phi} = 1$ enforces
\begin{align}
    \sum_{k = 0}^{2d^2-1} \bm{z}(\phi)_k^2 = 1,
\end{align}
where $\bm{z}(\phi)^{\mathrm{T}} = \big(\mathrm{Re}[\bm{\phi}_0], \mathrm{Im}[\bm{\phi}_0], \cdots, \mathrm{Re}[\bm{\phi}_{d^2 - 1}], \mathrm{Im}[\bm{\phi}_{d^2 - 1}])$. Thus, $\bm{z} \in \mathbbm{S}^{2 d^2 - 1}$. The reverse $\bm{z} \mapsto \ket{\phi(\bm{z})}$ map is constructed by straightforwardly reversing the steps above. Both maps, $\ket{\phi} \mapsto \bm{z}(\phi)$ and $\bm{z} \mapsto \ket{\phi(\bm{z})}$, preserve the Euclidean distance; in particular,
\begin{align} \label{Euc-to-Euc}
    \left\Vert \bm{z} - \bm{z}' \right\Vert_2 = \Vert \ket{\phi(\bm{z})} - \ket{\phi(\bm{z}')} \! \Vert_2 .
\end{align}

All in all, $\bm{z}$ being sampled from the Haar probability measure on $\mathbbm{S}^{2 d^2 - 1}$ exactly corresponds to the states
\begin{align*}
    \bm{\rho}(\bm{z}) \coloneq \T(\ket{\phi(\bm{z})})
\end{align*}
being sampled from the Hilbert--Schmidt probability measure on $\mathcal{H}_d$. Note that, when $\bm{z}$ runs through all of $\mathbbm{S}^{2 d^2 - 1}$, $\bm{\rho}(\bm{z})$ run through all of $\mathcal{H}_d$. Thus, for any real function of the state $f(\rho)$ [e.g., $\erg(\rho, H)$ or $S(\rho)$],
\begin{align} \label{behemoth}
\begin{split}
     &\prob\Big[ \big| f(\rho) - \avHS{f(\rho)} \big| > \ell \Big] =
     \\[2mm] 
     \prob\Big[& \big| f(\bm{\rho}(\bm{z})) - \bav{f(\bm{\rho}(\bm{z}))}e_{\text{Haar on} \; \mathbbm{S}^{2 d^2 - 1}} \big| > \ell \Big],
\end{split}
\end{align}
where the subscript HS in the first line means that the averaging is over the Hilbert--Schmidt probability measure on $\mathcal{H}_d$.

Furthermore, due to Eqs.~\eqref{erg_lips_euclid},~\eqref{hayden_ineq}, and~\eqref{Euc-to-Euc}, the functions $\erg(\bm{\rho}(\bm{z}), \, H)$ and $S(\bm{\rho}(\bm{z}))$ from $\mathbbm{S}^{2 d^2 - 1}$ to $\mathbbm{R}^1_+$ are Lipschitz continuous with respect to the Euclidean distance in $\mathbbm{R}^{2 d^2}$, with Lipschitz constants $L_{\mathrm{E}}^\erg$ and $L_{\mathrm{E}}^S$ given in Eqs.~\eqref{koxac1} and~\eqref{koxac2}. Therefore, Lévy's lemma applies to both of these functions. Using the bound in Eq.~\eqref{Levis} on $\erg(\bm{\rho}(\bm{z}), \, H)$ and $S(\bm{\rho}(\bm{z}))$ and invoking Eq.~\eqref{behemoth} at the end, we arrive at the concentration bounds in Eqs.~\eqref{conc_erg}--\eqref{width_ent} in the main text.

\section{Bures measure as quantum Jeffreys prior}
\label{app:bures}

The main reason the Bures measure is considered to be the analog of the Jeffreys prior in the quantum regime is that the Bures distance is Fisher-adjusted. Namely, for diagonal states, which are essentially classical probability distributions, the Bures distance coincides with the Fisher information metric \cite{Amari-Nagaoka_book_2000} (a.k.a. ``statistical distance'' in physics literature \cite{Wootters_1981, Braunstein_1994}). This metric is naturally related to statistical distinguishability and is invariant under sufficient statistic transformations \cite{Amari-Nagaoka_book_2000}. The Riemannian volume measure induced by the Fisher information metric is the Jeffreys prior. Thus, on the submanifold of diagonal states (in some basis), the Bures measure coincides with the Jeffreys prior. The isotropy (i.e., invariance under unitary rotations) of the Bures distance, and hence the measure, ensures that this is equally true in all bases.

The noninformativeness of the Bures measure is further corroborated by the fact that the Bures distance is Fubini--Study--adjusted and contractive. Indeed, the former means that, for pure states, the Bures distance coincides with the Fubini--Study metric, which is the most natural metric on pure states in that it is the only (up to a constant factor) Riemannian metric that is invariant under unitary transformations. As a result, the measure induced by it is Haar-uniform on pure states, and is thus maximally random. Finally, contractivity (or ``monotonicity'') \cite{Petz_1996, Bengtsson_book_2006} states that, for any completely positive trace-preserving (CPTP) map $\Phi$, $\DBu(\Phi(\rho), \, \Phi(\sigma)) \leq \DBu(\rho, \sigma)$, $\forall \rho, \sigma$. This means that the information encoded in the states $\rho$ and $\sigma$, as quantified by how distinguishable (i.e., far) they are from each other in Bures distance, can only decrease when we process these states. Thus, the Bures distance leaves no ``informational footprint'' on density matrices. This is, for example, contrasted by the Hilbert--Schmidt distance. While contractive under the action of unital channels \cite{Perez-Garcia_2006}, it can increase---thus falsely signaling information increase between the states---under the action of more general CPTP maps \cite{Ozawa_2000}.

The uniqueness of the Bures measure when imposing all the above desirables leaves it as the only choice (among measures induced by all other Fisher-adjusted Riemannian metrics) for a quantum generalization of the classical Jeffreys prior.


\bibliography{literature.bib}


\end{document}